\documentclass[11pt]{article} 
\usepackage{hyperref} 
\pdfoutput=1 
 
\usepackage{hyperref} 
\pdfoutput=1 
 
\begin{document} 
 
\title{What am I? \\ Supercooled droplet or ice?} 
 
\author{Carlo Antonini, Adrian Mularczyk, Tanmoy Maitra, \\ Manish K. Tiwari, Dimos Poulikakos \\ 
\\\vspace{6pt} Laboratory of Themodynamics in Emerging Technologies, \\
ETH Zurich, Sonneggstrasse 3, CH-8092, Switzerland} 
 
\maketitle 
 
 
\begin{abstract} 
In this fluid dynamics video we show the trick played by a supercooled
liquid water drop against a superhydrophobic surface. The water drop
shows a double personality, impacting onto the surface the first time while still in the liquid state,
and then re-impacting as a frozen ice crystal.
\end{abstract} 
 
 
\section{Introduction} 
 
The fluid dynamic video illustrates the various interactions between a
liquid water drop and a superhydrophobic surface. As well-known,
a water drop will typically rebound from a superhydrophobic surface after impact (first impact),
thanks to a special wetting conditions, i.e. the so-called Cassie-Baxter state. \\ 
However, if the impact speed reaches a value higher than the critical speed,
the drop or a part of it may remains attached to the surface, unable to rebound.
This is due to the wetting transition from the Cassie-Baxter to the Wenzel wetting state (second impact).
Both this phenomenon can be observed at room temperature.\\

Interesting phenomena, however, can be observed in an environment where
the temperature falls below water freezing temperature, i.e. 0$^\circ$C.
In the third and last impact, the environment, the drop and the surface are all at -16$^\circ$C.
As common sense would suggest, water should be frozen. However, ultra-pure liquid water
can be kept supercooled without freezing. At -16$^\circ$C, the water drop is heavily
supercooled but still liquid. Interestingly, the supercooled
liquid drop initially impacts on the superhydrophobic surface and bounces off,
as for the impact case at room temperature. However, during the post-rebound flight,
the drop freezes in mid-air, and re-impacts onto the surface for the second time
in the form of an ice crystal, this time unable to rebound.

 
Two sample videos are 
\href{}{Video 1} and 
\href{}{Video 2}.

\end{document}